\documentclass[pdf]{iucr}
\journalcode{S} 

\usepackage{graphicx}

\usepackage{url}

\usepackage{xcolor}

\usepackage{amsmath}
\usepackage{amssymb}
\usepackage{lineno}
\usepackage{array}
\usepackage{algorithm}
\usepackage{algpseudocode}

\newcommand{\un}[1]{\ensuremath{\,\mathrm{#1}}}

\begin{document}                  

\title{darfix: Data analysis for dark-field X-ray microscopy}
\shorttitle{darfix}

\cauthor[a]{J{\'u}lia}{Garriga Ferrer}{julia.garriga@esrf.fr} {} 
\author[a]{Raquel}{Rodr{\'\i}guez-Lamas}
\author[a]{Henri}{Payno}
\author[a]{Wout}{De Nolf}
\author[a]{Phil}{Cook}
\author[a]{Vicente Armando}{Sol{\'e} Jover}
\author[a]{Vincent}{Favre-Nicolin}
\author[a]{Can}{Yildirim}
\author[a]{Carsten}{Detlefs}

\aff[a]{ESRF, The European Synchrotron, 71 Avenue des Martyrs, CS40220, 38043 Grenoble Cedex 9, \country{France}}

\aff[b]{Department of Physics, Technical University of Denmark, 
2800 Kgs.~Lyngby, \country{Denmark}}

\shortauthor{J. Garriga \textit{et al.}}


\maketitle                        

\begin{synopsis}
 \end{synopsis}

\begin{abstract}

A Python package for the analysis of dark-field X-ray microscopy (DFXM) and rocking curve imaging (RCI) data is presented. 
\textit{darfix} provides a set of data processing and visualization tools that can be either imported as library components or accessed through a graphical user interface (GUI) as an Orange add-on.  
In the latter case, the different analysis modules can be easily chained to define computational workflows.
Operations on larger-than-memory image sets are supported through the implementation of online versions of the data processing algorithms, effectively trading performance for feasibility when the computing resources are limited. The software can automatically extract the relevant instrument angle settings from the input files metadata. The currently available input file format is EDF and in future releases HDF5 will be incorporated.


\end{abstract}


\section{Introduction}

Dark-field X-ray microscopy (DFXM) is a novel full-field imaging technique that non-destructively maps the 3D structure, orientation and strain of deeply embedded crystalline elements, such as grains or domains \cite{Simons2015, Poulsen2017, Poulsen2020, Yildirim2020}. 
Direct-space images are formed by placing an X-ray objective lens along the diffracted beam, affording a spatial resolution on the order of 100\un{nm}, while maintaining a working distance between the sample and X-ray objective lens that is in the cm-range.  
The first implementation of a dedicated dark-field x-ray microscope was recently installed on beamline ID06-HXM of the European Synchrotron Radiation Facility \cite{Kutsal2019}.  
Since its installation, this instrument has been used to investigate a variety of scientific subjects, including the domain evolution in ferroelectrics \cite{Simons2018}, the austenitic transformation in shape memory alloys \cite{Bucsek2019}, recovery in metals \cite{Mavrikakis2019, Ahl2020}, embedded particles in steel \cite{Hlusko2020},  visualization of dislocation structures \cite{Jakobsen2019,Dresselhaus2020}, and the structure of biominerals \cite{Cook2018,Schoeppler2022}. 

DFXM is conceptually similar to dark-field electron microscopy in transmission electron microscopy (TEM), which is used to selectively image strain and orientation across materials science, physics, geoscience and numerous other fields \cite{Williams1996, Nellist2000, Morones2005}.


\section{Dark Field X-ray Microscopy}

\begin{figure}
    \begin{center}
    \resizebox{0.9\columnwidth}{!}{\includegraphics{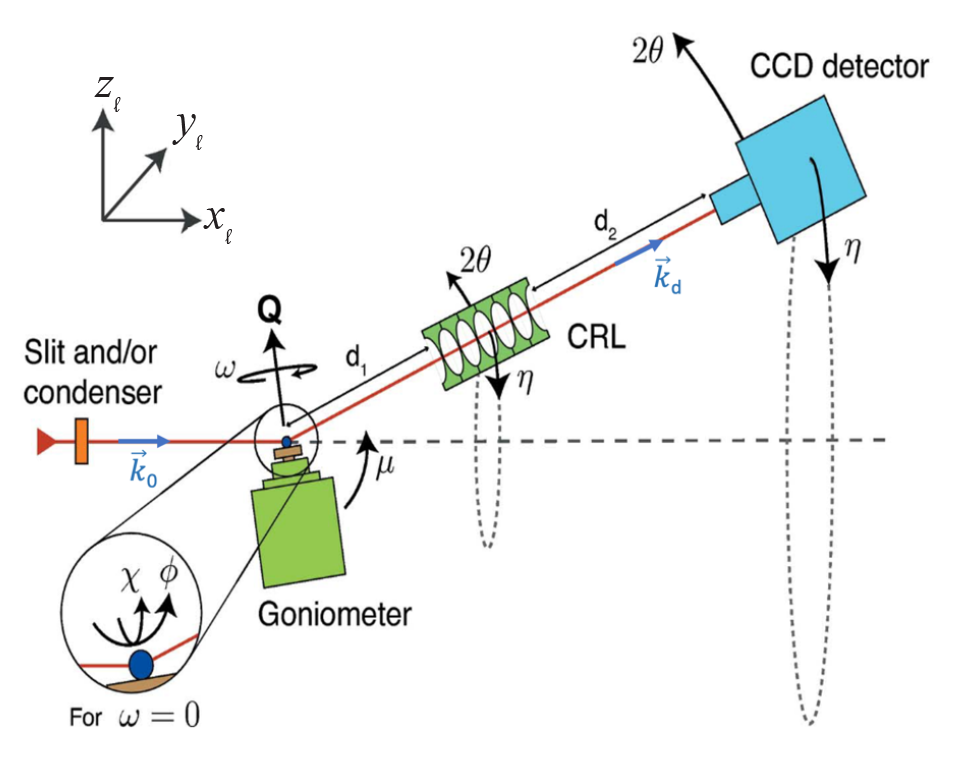}}
    \end{center}
\caption{Geometry of the dark-field X-ray microscope at ID06-HXM at the ESRF. The incident beam $\vec{k}_0$ travels along the laboratory $x_\ell$ axis. The optical axis of the objective (CRL) lens is aligned to the direction of the diffracted beam $\vec{k}_d$. The pivot point of the goniometer and sample is coincident with the intersection of these two optical axes. Vector $\vec{Q}$ defines the local scattering vector at a given point $\vec{r}(x,y,z)$ within the sample, and may be parameterized by the scattering angle $2\theta$, the azimuthal angle $\eta$ and the length of the vector $\left|\vec{Q}\right|$.  The value of $\left|\vec{Q}\right|$ is related to the spacing of the lattice plane  being measured, $d_{hk\ell}$, and the X-ray wavelength, $\lambda$, by Bragg's law. The goniometer is associated with a base tilt, $\mu$, an $\omega$-rotation around $\vec{Q}$ and two tilts, $\chi$ and $\phi$. $d_1$ is the distance from the sample to the entry point of the objective and $d_2$ is the distance from the exit point of the objective to the detector. The positive directions of the angles are indicted. This figure is adapted from \citeasnoun{Poulsen2017}.
}

\label{figure-s1}
\end{figure}


The geometry of DFXM is illustrated in Fig.~\ref{figure-s1}. Further details can be found in \citeasnoun{Poulsen2017} and 
\citeasnoun{Kutsal2019}. 
A nearly monochromatic and nearly collimated X-ray beam illuminates the sample. 
This beam may be condensed in the vertical and horizontal direction to increase the beam intensity in the field of view, or to selectively illuminate a thin layer within the sample. 
The energy bandwidth is of the order $\Delta E/E \approx 1.4\cdot 10^{-4}$. 

The goniometer is designed to access diffraction angles in a nearly vertical scattering geometry, and probe reciprocal space only in the immediate vicinity of a given Bragg reflection $(h,k,\ell)$. 
The current implementation of DFXM at ID06-HXM, ESRF achieves this by moving the sample along a combination of $\mu, \omega, \chi$ and $\phi$ rotation stages, see Fig.~\ref{figure-s1}.  
The direction of the diffracted beam is characterized by the scattering angle, $2\theta$, and the azimuthal angle, $\eta$. In most experiments the sample is aligned such that $\eta \approx 0$.

The optical axis of an X-ray objective is aligned to the diffracted beam to produce a magnified image (inverted in both directions) on the 2D detector. 
The magnification can be calculated from the distances $d_1$ (sample to objective) and $d_2$ (objective to detector), and experimentally verified by observing the displacement of the image upon small calibrated translations of the sample.

In addition to the magnified DFXM images, there are further 2D detectors to record non-magnified images, e.g. a ``near field camera'' positioned directly downstream of the sample \cite{Kutsal2019}. This can be used for classical diffraction topography and its extension, rocking curve imaging \cite{Tran2017}.


\subsection{Scan types}

The 2D images recorded as function of the different rotation angles form the main data sets of DFXM. \textit{darfix} facilitates the treatment of the individual raw images and the systematic analysis of scans.

Scan ranges and step size are generally a compromise of completeness and resolution vs.~scan duration.

The following recurring scan types are commonly used in DFXM:


\subsubsection{Rocking curve imaging:}


\label{sec_rci}
Rocking curve imaging is an extension of classical diffraction topography: The diffraction topograph is measured as a function of the ``rocking'' angle, $\mu$ (see Fig.~\ref{figure-s1}). Rocking curve imaging is typically performed without the magnifying objective lens, using the near field camera. In this case the resolution in the $2\theta$ and transverse ($\chi$-) directions is relatively low \cite{Tran2017}.
It can, however, also be carried out in DFXM mode.

The rocking curve imaging module provides functions similar to the RCIA code \cite{Tran2017} within the workflow of \textit{darfix}. 
Its main function is to analyze the rocking curve of each pixel by fitting to a peak shape, e.g.~a Gaussian.
After the fit,  maps of the fit parameters (constant background, integrated intensity, peak position and peak width) as function of pixel position are generated \cite{Tran2017}, as shown in Fig.~\ref{figure-s5}. 

As a computationally less intensive alternative to fitting, moments of the intensity as function of the angle can be computed. 
For a perfectly Gaussian distribution, the zero order moment corresponds to the integrated intensity, the first order moment (center of mass) to the peak position, and the second order moment (variance) to the square of the rms peak width.
These can be used as starting values for the Gaussian fit. 
\textit{darfix} can furthermore calculate the third (skewness) and fourth (kurtosis) moments.
The moments can also be presented as color maps.

\subsubsection{Mosaicity scans:}

Mosaicity scans can be seen as a generalization of rocking curve imaging, taking advantage of the improved resolution in the transverse ($\chi$-) direction due to the limited angular acceptance of the objective lens \cite{Poulsen2017}. 
Similar to rocking curve imaging, the dependence of the intensity on two rotation angles, the ``rocking'' and ``rolling'' angles, $\mu$ and $\chi$ respectively, is analyzed for each pixel.
Typically moments are calculated. 
The first-order moments (center-of-mass) correspond to the local orientation of the diffracting planes.
Due to the narrow acceptance of the objective lens in the $2\theta$ direction, the magnitude of the scattering vector remains constant \cite{Poulsen2017} such that only shear strains and lattice rotations are measured \cite{Poulsen2021}.
Again, the results can be represented in a color map, using a 2D color plane that encodes both rotation angles.
Some examples are shown in Fig.~\ref{figure-s2} (section~\ref{section-mosa}), where the center of mass color maps for $\mu$ (Fig.~\ref{figure-s2}a) and $\chi$ (Fig.~\ref{figure-s2}b) are shown.


\subsubsection{Strain scans:}

In strain scans, the scattering angle $2\theta$ is varied in order to probe spatial variations in the $d$-spacing of selected Bragg reflection \cite{Poulsen2017,Poulsen2021}. 
Typically strain maps are constructed from  a series of rocking scans performed consecutively over a given range of scattering angles. 

\subsubsection{Combined mosaicity-strain scans:}

By recording a (2D) mosaicity map instead of a (1D) rocking curve scan at each $2\theta$ position, all 3 directions in reciprocal space are probed.
This is the most complete type of DFXM scan.
However, as a 3D mesh of motor positions must be scanned, this scan type also requires the longest data acquisition time and yields the largest raw data volumes.

\subsubsection{Reciprocal space mapping:}

This type of scan is similar to rocking curve imaging, but without the objective lens. The images are recorded on a large field of view camera positioned in the farfield \citeaffixed{Kutsal2019}{i.e.~${
\approx}5\un{m}$ from the sample, }. This type of scans provides angular information in three dimensions, $\mu$ (scan axis), $\eta$ and $2\theta$ (derived from the pixel position, see Fig.~\ref{figure-s1} and \citeasnoun{Poulsen2017}). 
Similar to rocking curve real space imaging, the analysis includes integrated intensity and center-of-mass visualization. 
Reciprocal space mapping can be used, e.g., to determine twin relationships between domains in ferroelastic materials \cite{Gorfman2022}.

\subsubsection{Layer scans:}

All of the scans listed above can be performed with a line-focused beam illuminating a single, ${\approx}150\un{\mu m}$ thick layer within the sample.
Layers recorded at a series of heights within the sample can then be combined into 3D volume maps. 


\section{The \textit{darfix} codebase}

\textit{darfix} is a Python library that provides a set of computer vision techniques for the analysis of dark-field X-ray microscopy data. 
It is split into two modules: 
a back end that contains the core processing utilities, algorithm implementations and abstractions, and a graphical user interface (GUI). 
The typical user would interact with the core module through the GUI, but it is possible and straightforward to access the core methods directly by importing that part of the library. 
The central code object is \texttt{Dataset}, which encapsulates the properties of a stack of images (multiple rotation angles can vary within a single stack of images, each axis adds a dimension to the image stack).
Operations on the source data (e.g.~blind source separation, hot pixel removal, shift correction) are implemented as pure transformations acting on this entity, such that there are no side effects and the operations can be chained together to define a computational workflow. 
Furthermore, as datasets are commonly of the order of several hundred gigabytes, the software provides online versions for many of the data processing algorithms: the output is build incrementally while adapting the changes to the input. These options work without loading the data in the computer memory so that computation is feasible on memory constrained machines. If the option \textit{data in disk} is selected when loading the data, these versions will be automatically used along the workflow.

\textit{darfix} does not have a stand-alone application, but includes an Orange \cite{orange} add-on for the definition of workflows. Orange is a platform to perform data analysis and visualization, and which allows for the creation of workflows using the \textit{darfix} GUI.
A typical workflow comprises task such as raw data selection, background removal, region-of-interest selection, scan variable identification, scan processing (e.g.~rocking curve fitting), display of the results, etc. (see Fig.~\ref{figure-s3}).
All the tasks in \textit{darfix} are independent (in the GUI as well as in the processing part), so that for every task there is a different widget. 
These widgets are based on Qt \cite{Qt} and silx \cite{silx}. 
They can be linked through the workflow created in Orange. 
In addition, workflows in Orange can be called from the command line with different input datasets.
The algorithms that were used in the original workflow will be then reproduced without the need of the GUI. 
In addition, a workflow defined from the GUI can be saved with it settings and reused using \textit{ewoks} \cite{ewoks}. 
This allows users to launch the same workflow (algorithms and settings) but with a different input data set.
This is particularly useful when the same workflow has to be carried out for a series of datasets, e.g.~the same analysis has to be applied to every layer in a 3D volume map.

\textit{darfix} contains a series of image processing methods for the analysis of DFXM scans. These methods can be separated in three parts: the selection of the data, the pre-processing algorithms, and the analysis and visualization of the results. 
The following sections will show the used techniques and how they are implemented.


\subsection{File inputs: data selection}

Currently, \textit{darfix} accepts the ESRF data format (.edf) of the ID06 detectors, and is able to automatically characterize the relevant instrument angle settings by analyzing the input files' metadata. HDF5 \cite{hdf5, h5py} is accepted, but only without dimension definition as the analysis of metadata is still under development.

Output data can be saved from the GUI in different formats: EDF, HDF5 \citeaffixed{NXdata}{NeXus data format, }
, CSV, TIFF, Numpy, ASCII and PNG. In the analysis there is usually the possibility to export specific data in a HDF5 file, which can then be used for any needed post-processing.
Alternatively, when working without the GUI, \textit{numpy} arrays can be extracted from the \texttt{Dataset} object for further processing.

Modified data is automatically saved on disk under a treated data folder chosen by the user. By default, only the most recent data of the workflow is saved, but there exists the possibility to copy the data using the \textit{Data copy} widget.

Exporting data in a format compatible with 3D visualization and processing software, e.g. Paraview \cite{Paraview1,Paraview2} is planned for a future release.


\subsection{Pre-processing}

\subsubsection{Noise removal:}

It is common to have noise along the set of data images, that either comes from the lens, the environment or the diffraction of the sample.
The first step (or second if we apply a region of interest to the data) of the pre-analysis is to detect and remove this noise from the sequence of images. \textit{darfix} provides the following tools for noise removal:
\paragraph{Background subtraction:} This is a widely used approach that calculates the foreground mask of an image by subtracting it from a background model containing the static part of an image sequence. In \textit{darfix}, the background is an image that can either be calculated using the mean or the median of a set of data images (for example dark images from the scan). This background image is then subtracted from each of the original images to obtain the foreground mask. While both mean and median filters can be used depending on the input dataset, the median filter is more robust and realistic (the output pixels are always one of the stack) than the mean filter, although the median filter needs either memory or time. To speed up the process when the data is not in memory the user has two options:
\begin{itemize}
    \item Select chunks of a certain shape so that the median is applied consecutively in all of them. Although this methods obtains a median of all the images, it is still time consuming as the input/output operations slow a lot the process.
    \item Only use a subgroup of the stack of images to calculate the median: the user inputs a step $k$ value so that the algorithm selects the images with index $i, i+k, i+2k,...$.
\end{itemize}
\paragraph{Hot pixel removal:} Sometimes, after performing background subtraction on the images, isolated groups of pixels appear at some or all of the images in the stack. These pixels are called ``hot'' because they have higher intensity than the other pixels around them, and they are usually not part of the object we want to analyze but noise that needs to be retrieved.
As the hot pixels are independent from one image to the other, this technique can be applied sequentially to the stack:
\begin{itemize}
    \item Apply a median filter to every image of the stack and subtract it from the original image.
    \item The hot pixels are identified as the ones with higher value than the standard deviation of this subtraction.
    \item The hot pixels intensity values are replaced by the values of their corresponding pixels in the image with the median filter applied.
\end{itemize}
\paragraph{Threshold removal:} Pixels with values lower than specified will be set to 0.

The use of any or all of the noise removal tools is optional.

\subsubsection{Image registration:}

Alignment errors in the experimental hardware frequently cause a shift of the image as function of rotation angle. 
\textit{darfix} contains a module for the detection and correction of such shifts. 
We assume that the shift is a linear function of rotation angle, i.e. $\mathbf{v}(\alpha) = \alpha \mathbf{v}_0$, where $\alpha$ is the rotation angle relative to the scan center, and $\mathbf{v}_0$ is the displacement vector in pixels/degree.
The algorithm used to find the shift, which has proven to give acceptable results in a considerably fast manner, is the following:

\begin{itemize}
\item Two images are obtained from the sum over axis 0 of the first and second half of the dataset.
\item The shift vector $\mathbf{v}'$ between the two resulting images is determined using the \texttt{scikit-image} \cite{scikit-image} function \textit{registration.phase\textunderscore cross\textunderscore correlation}.
\item The linear shift is in the direction of the normalized shift: $\mathbf{v}_0 = h \,\mathbf{v}'/\left|\mathbf{v}'\right|$.
\item The scaling factor $h$ remains to be determined -- it depends on how the intensity of the images varies as function of $\alpha$:
\begin{algorithmic}
\State $\mathbf{v}_0^* = \dfrac{2\mathbf{v}_0}{num\_images}$;  $\varepsilon = 3{\lVert \mathbf{v}_0^* \rVert }_{2}$
\For{$h^* \gets 0$ to $3\varepsilon$}
\For{$i \gets 0$ to $num\_images$}
\State $\mathbf{v}^*(i) = ih^*\mathbf{v}_{0}$
\State Shift image i using $\mathbf{v}^*(\alpha)$ via an affine transformation
\EndFor
\State Compute score using normalized variance of the images sum
\EndFor
\State $h$ is the $h^*$ with maximum score
\end{algorithmic}
\end{itemize}

Once $\mathbf{v}$ has been determined, the images are shifted with subpixel accuracy using the Discrete Fourier Transform (DFT) algorithm.
The library OpenCV \cite{opencv_library} is used both for the affine transformations and for the DFT algorithm.

One could use other strategies, e.g.~a non-linear fit, to optimize $h$.
In some cases, e.g.~when summing the uncorrected images leads to blurring, better results are obtained when the shift correction is run several times.

It is possible to find and apply the shift along a chosen dimension. 
In this case, a different shift is detected for every value of the chosen dimension. 
These shifts can then be applied to their corresponding images.


\subsection{Scan analysis}

After pre-processing, data can be further analysed according to the type of scan performed during the experiment, see above.

\subsubsection{Rocking curves fitting:}
\label{sec_rci2}

As described in section \ref{sec_rci}, \textit{darfix} allows for a rocking curve analysis.

The intensity of a pixel, $I_{\mathrm{meas},xy}$, as function of a motor position, typically the rocking angle $\mu$, is fitted to a Gaussian,
\begin{equation}
    I_{\mathrm{fit},xy}(\mu)
    =
    b_{xy} + A_{xy} \cdot
    \exp\left[-\frac{1}{2} \frac{(\mu - p_{xy})^2}{\sigma_{xy}^2} \right].
\end{equation}


Fitting all pixels in this way results in maps of the background ($b_{xy}$), amplitude ($A_{xy}$), peak position $(p_{xy}$) and peak width ($\sigma_{xy}$, full width at half maximum (FWHM) $= 2.355 \sigma_{xy}$). 
Furthermore the fit generates a map of the chi-squared values,
\begin{equation}
    \chi^2_{xy} = \sum_{\mathrm{images}}
    \left( I_{\mathrm{fit},xy} - I_{\mathrm{meas},xy} \right)^2,
\end{equation}
that can be used to assess deviations from a simple Gaussian profile.

For faster results, pixels with low intensities can be omitted from the fit.

2-dimensional scans, where e.g.~two motors $\mu$ and $\chi$ are varied, can be fitted to a bivariate Gaussian, 

\begin{eqnarray}
    \lefteqn{I_{\mathrm{fit},xy}(\mu,\chi)
     = 
    b_{xy}
    +
    A_{xy}
    \cdot
    \exp\left[
    -\frac{1}{2 (1-\rho_{xy}^2)}
    \left(
        \left(
            \frac{\mu - p_{\mu,xy}}{\sigma_{\mu,xy}}
        \right)^2
\right.\right. } \nonumber \\ & & \left. \left.
        -2 \rho_{xy}
        \left(
            \frac{\mu - p_{\mu,xy}}{\sigma_{\mu,xy}}
        \right)
        \left(
            \frac{\chi - p_{\chi,xy}}{\sigma_{\chi,xy}}
        \right)
        +
        \left(
            \frac{\chi - p_{\chi,xy}}{\sigma_{\chi,xy}}
        \right)^2
    \right)
    \right].\ \ \ \ \ 
\end{eqnarray}

Here $p_{\mu,xy}$ and $\sigma_{\mu,xy}$ are maps of the peak position and peak width of motor $\mu$, $p_{\chi,xy}$ and $\sigma_{\chi,xy}$ are maps of the peak position and peak width of motor $\chi$, and $\rho_{xy}$ is the map of the Pearson correlation coefficient.

Additional line shapes are planned for a future release.

\subsubsection{Grainplot:}

For scans where more than one motor varies, such as mosaicity ($\mu$--$\chi$ scans), the Grainplot tool can calculate moments, e.g. the integrated intensity, center-of-mass and standard deviation. 
In particular, for mosaicity scans, the center-of-mass (COM) of the two scan motors are interpreted as components of a 2D vector in the color plane. Both components can then be displayed in a single graph. Intensity contours in the corresponding color map represent a local pole figure of the volume of interest.

\subsubsection{Blind source separation:}

Blind source separation (BSS) comprises all techniques that try to decouple a set of source signals from a set of mixed signals with unknown (or very little) information \cite{Herault1985}. Depending on the assumptions of the data, different BSS techniques can be used.

In DFXM, diffracting elements such as (sub)grains or ferroelastic domains can be interpreted as source signals that contribute to the images.
BSS can then be used to identify these elements and extract the corresponding rocking curves, reciprocal space maps, etc. in a given dataset.

\textit{darfix} implements several BSS techniques.
For an easier analysis of the data, the images are flattened and set as rows of a single matrix $\mathbf{X}$. The goal is to find two matrices, $\mathbf{H}$ and $\mathbf{W}$, which product approximates $\mathbf{X}$, and so that $\mathbf{H}$ rows are the components we are looking for. $\mathbf{H}$ is called the components matrix and $\mathbf{W}$ the mixing matrix. This approximation is done by optimizing the distance \textit{d} between $\mathbf{X}$ and the matrix product $\mathbf{WH}$. This approximation is usually done using the Frobenius norm.

The BSS techniques implemented in \textit{darfix} are:
\begin{itemize}

\item Principal Component Analysis (PCA): PCA is a BSS technique to recover a collection of orthogonal vectors $\mathbf{H}$, called principal components, and perform a change of basis on the data which usually only uses the principal components that better represent the data.

If the data is in memory, a PCA implementation based on \citeasnoun{MARTINSSON2011} and \citeasnoun{Tipping1999} is included. Otherwise we use the incremental PCA model from \citeasnoun{Ross2008}. In both cases the \texttt{scikit-learn} implementation \cite{scikit-learn} is used.

As the principal components are orthogonal, they do not satisfy the non-negativity criterion for image intensity.
The principal components therefore can not be interpreted as diffracting crystal elements.

The eigenvalues of the principal components are very useful to know how each of the components is present in the dataset and to approximate a number of needed components to represent it. This number of components can then be used as input parameter for the other BSS methods described below.


\item Non-negative Independent Component Analysis (NICA):
working with images, we can assume a constraint that can help in the separation of the components: the non-negativity of the sources.
NICA finds the components by assuming that they are non-Gaussian and statistically independent from each other (a technique known as independent component analysis, ICA,  \cite{hyvarinen2001}), while restricting $\mathbf{H}$ to be non-negative. 
This alternate way of approaching ICA has several algorithms and methods presented such as \cite{yuan2004} and \cite{ouedraogo2010}. \textit{darfix} implements the method described in \cite{Oja2004}. However, NICA does not require the mixing coefficients to be non-negative.
Negative contributions of diffracting elements to the image intensity are unphysical.

\item Non-negative Matrix Factorization (NMF):
we can constraint both the sources and the mixing elements ($\mathbf{H}$ and $\mathbf{W}$) to be non-negative. 
This constraint is called non-negative matrix factorization and is another widely-used technique for solving BSS \cite{Lee1999}. \textit{darfix} uses the implementation in \texttt{scikit-learn} to compute NMF when the data is in memory. Otherwise our own implementation is used based on the multiplicative update rule \cite{Lee2001}. This method is applied in chunks to avoid having all the data in memory.

\item (NICA + NMF): The non-uniqueness (non-convexity) property of NMF implies that the solution depends on the initial factor matrices. To solve this problem we implement the idea presented in \cite{Kitamura2016} which suggests that a good initialization is based on the factorization given by non-negative ICA.
\end{itemize}


\subsection{Graphical user interface}

\onecolumn

\begin{figure}
    \begin{center}
    \resizebox{0.9\columnwidth}{!}{\includegraphics{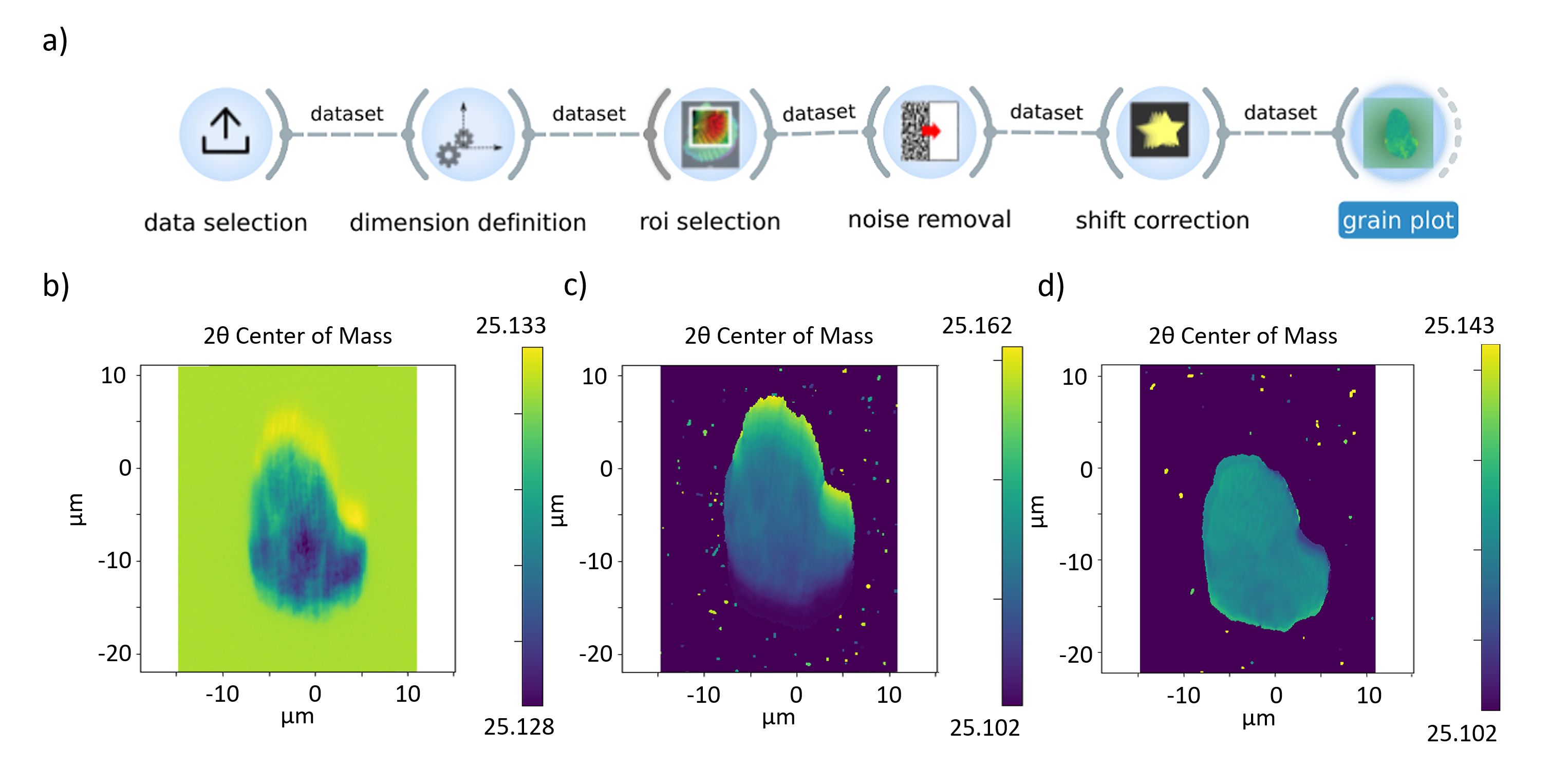}}
    \end{center}
    \caption{a) \textit{darfix} GUI where a data preprocessing thread is shown, finalised by the Grainplot widget. b) raw data plotted as COM c) after background removal and threshold removal d) after hot pixel removal and shift correction. The color bars represent scattering angles ($2\theta$).
    }
\label{figure-s3}
\end{figure}

\twocolumn

The GUI is programmed in Qt \cite{Qt} and silx \cite{silx} and it is structured to have a different widget for each step in the data processing workflow.
Orange \cite{orange} is used to link all these widgets into a single workflow and to pass information between them. Every widget returns a new \texttt{Dataset} object which is the input to the next step of the workflow.
Fig.~\ref{figure-s3}a shows an example of a typical workflow comprising data selection, different pre-processing steps, and shift correction. Fig.~\ref{figure-s3}b, c and d show examples of Grainplot of the intermediate and final results of the workflow (see below for details).


\subsection{Open source, documentation and tutorials}

\textit{darfix} is open source under MIT license. The software can be installed using \texttt{pip}, see \url{https://pypi.org/project/darfix/}.
User documentation can be found at \url{https://gitlab.esrf.fr/XRD/darfix/-/blob/master/docs/source/tutorials/darfix_guide.pdf}.


\section{Examples}

Fig.~\ref{figure-s3}a shows an example of a typical workflow comprising data selection, different pre-processing steps, and shift correction. Fig.~\ref{figure-s3}b, c and d show examples, plotted as a Grainplot COM, of the intermediate and final results of the workflow (see below for details). The represented data corresponds to a (200) reflection of a grain in a Fe-3$\%$Si sample that was produced as explained in \citeasnoun{Mavrikakis2019}.

\subsection{Data selection and dimension definition}

The first step of the workflow is to select the input data. Next, the dimensions are defined, i.e. the motor(s) varying during the scan and the number of points along each scan axis is determined (see Fig.~\ref{figure-s3}). 
\textit{darfix} will attempt an automatic dimension definition. This will result in the automatic identification of the moving motors, the angular ranges, the step and the number of steps on a scan.
At the current version, dimension definition is only available for data in EDF format, where motor positions are automatically extracted from the metadata. Data in HDF5 can be analyzed, but at present the metadata is not used.
After, a region of interest can be defined in order to reduce the data volume and speed up processing.

\subsection{Noise removal and Shift correction}

Next, noise removal and shift correction, as detailed above, are applied. The effect of these steps on the data is illustrated in Fig.~\ref{figure-s3}.
Fig.~\ref{figure-s3}b shows the raw data, followed by Fig.~\ref{figure-s3}c where a threshold removal had been executed and Fig.~\ref{figure-s3}d where hot pixel removal was also applied.

\subsection{Rocking curve imaging}
\begin{figure}
    \begin{center}
    \resizebox{0.75\columnwidth}{!}{\includegraphics{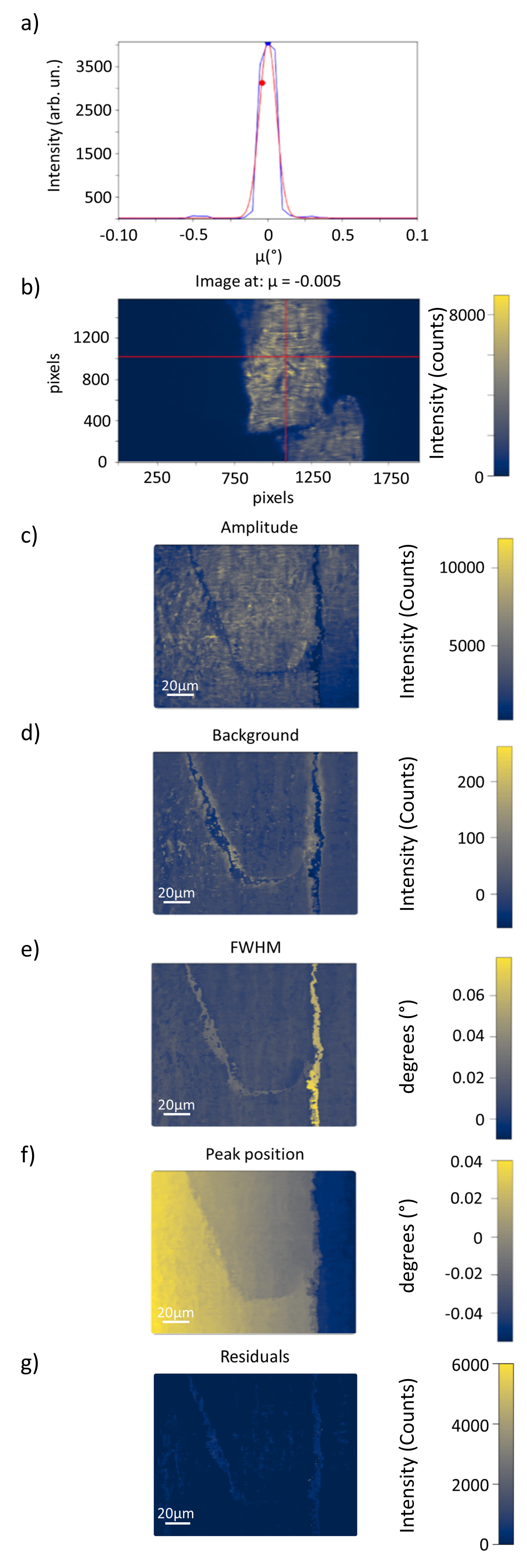}}
    \end{center}
\caption{Example of 1D rocking curve analysis, (a) fit of the curve corresponding to pixel indicated in (b) image at $\mu=-0.005\deg$ with a cross marking the pixel of interest. Maps of the full stack of images composing the rocking scan, generated by the pixel by pixel fit of (c) amplitude (d) background (e) full with half maximum (f) peak position (g) residuals.
}
\label{figure-s5}
\end{figure}

The rocking curve widget performs the fit as described in sections \ref{sec_rci} and \ref{sec_rci2} at each pixel of the data obtained from a rocking scan. 
Moreover, \textit{darfix} allows to input a 2D scan into the rocking curve widget and performs a 2D fit of the  mosaicity data (where the moving motors correspond to angular motions in $\mu$ and $\chi$).

Fig.~\ref{figure-s5} shows an example of the maps obtained from the rocking curve analysis of the (200) reflection of Al \cite{LeoraPrivate}, as well as the local rocking curve of a selected pixel. 
While Fig.~\ref{figure-s5}b presents the image obtained at a certain $\mu$ the fit maps Fig.~\ref{figure-s5} c-g provide information extracted from the fit of each pixel over the whole rocking curve. 
In this example it can be seen that the ``boundaries'' between two regions that diffract at different $\mu$ angle are the hardest to fit, as shown by the residuals  ($\chi^2$) maps. 
These boundaries show larger FWHM possibly related to a slightly gradual change in orientation. The peak position map shows a constant gradient, that can be related to an homogeneous deformation of the crystal. 



\subsection{Mosaicity and strain scan}

\label{section-mosa}

\begin{figure}
    \begin{center}
    \resizebox{0.9\columnwidth}{!}{\includegraphics{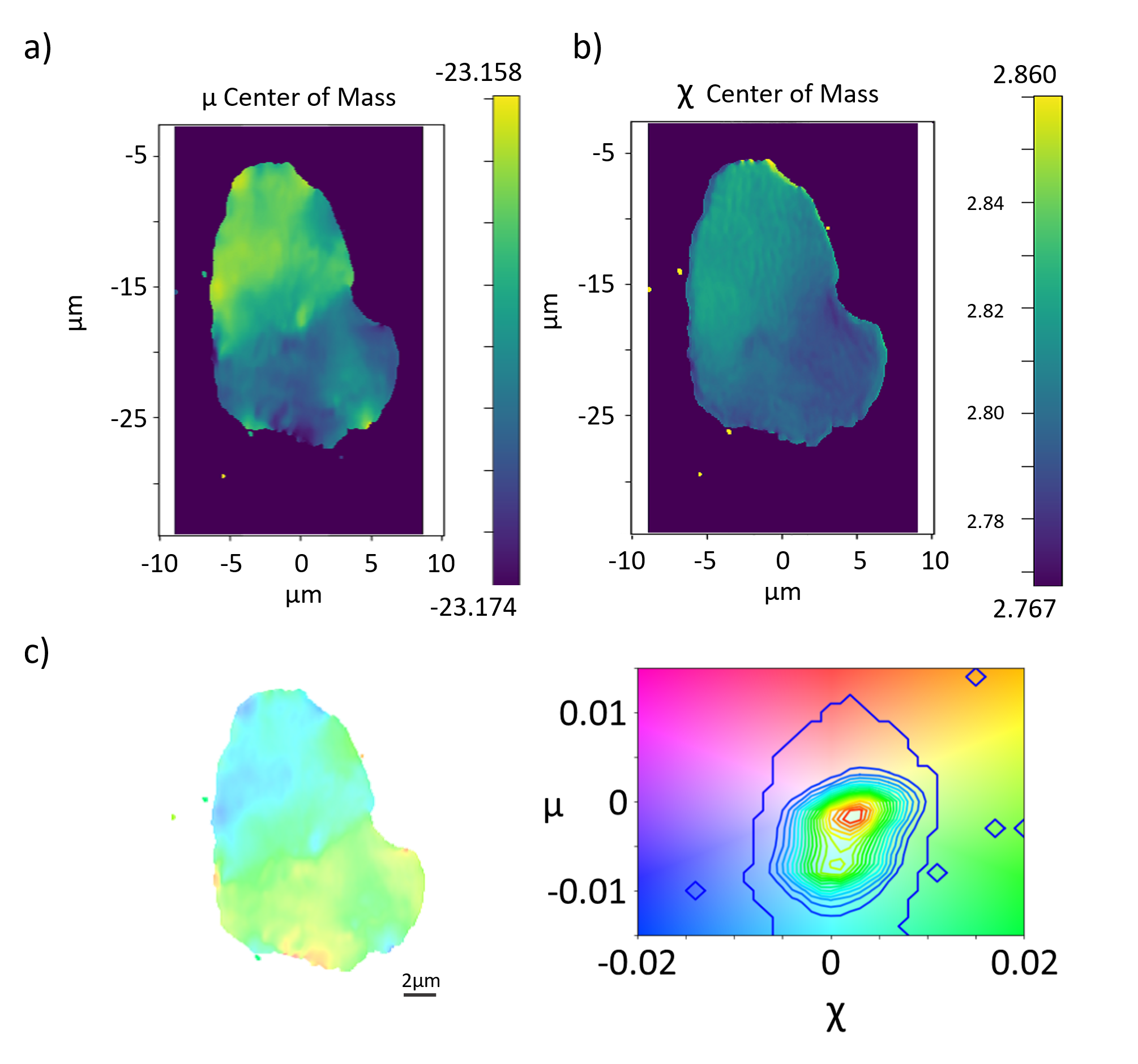}}
    \end{center}
\caption{Center of mass maps a) for $\mu$ and b) $\chi$ c) Mosaicity map and orientation distribution color key of the mosaicity map with an overlaid contour map of the integrated intensity indicates the angular spread for $\mu$ and $\chi$.
}
\label{figure-s2}
\end{figure}

Figure \ref{figure-s2} shows a mosaicity map obtained from the projection of a grain in Fe-3$\%$Si \cite{Mavrikakis2019}, in this case the (200) reflection. 
Fig.~\ref{figure-s2}a shows the peak position map of the pitch (rocking angle $\mu$), whereas Fig.~\ref{figure-s2}b shows the peak position in roll (angle $\chi$). 
The two center-of-mass angles can be combined into a color vector (Fig.~\ref{figure-s2}c). 
This type of maps represent the local crystallographic orientation around the chosen Bragg reflection. 
The contours in the color key (Fig.~\ref{figure-s2}d) thus represent a local pole figure of the (200) reflection.

Strain scans (varying $\mu$ and $2\theta$) can be analyzed in the same way. The interpretation, however, is different, as these scans measure the relative axial strain along a given crystallographic plane. In contrast to mosaicity scans where the lattice distortion and orientation are measured, strain scans provide information about the variation of the \textit{d}-spacing of a given \textit{hkl} plane of a crystal.  

\subsection{Blind source separation}

\begin{figure}
    \begin{center}
    \resizebox{0.9\columnwidth}{!}{\includegraphics{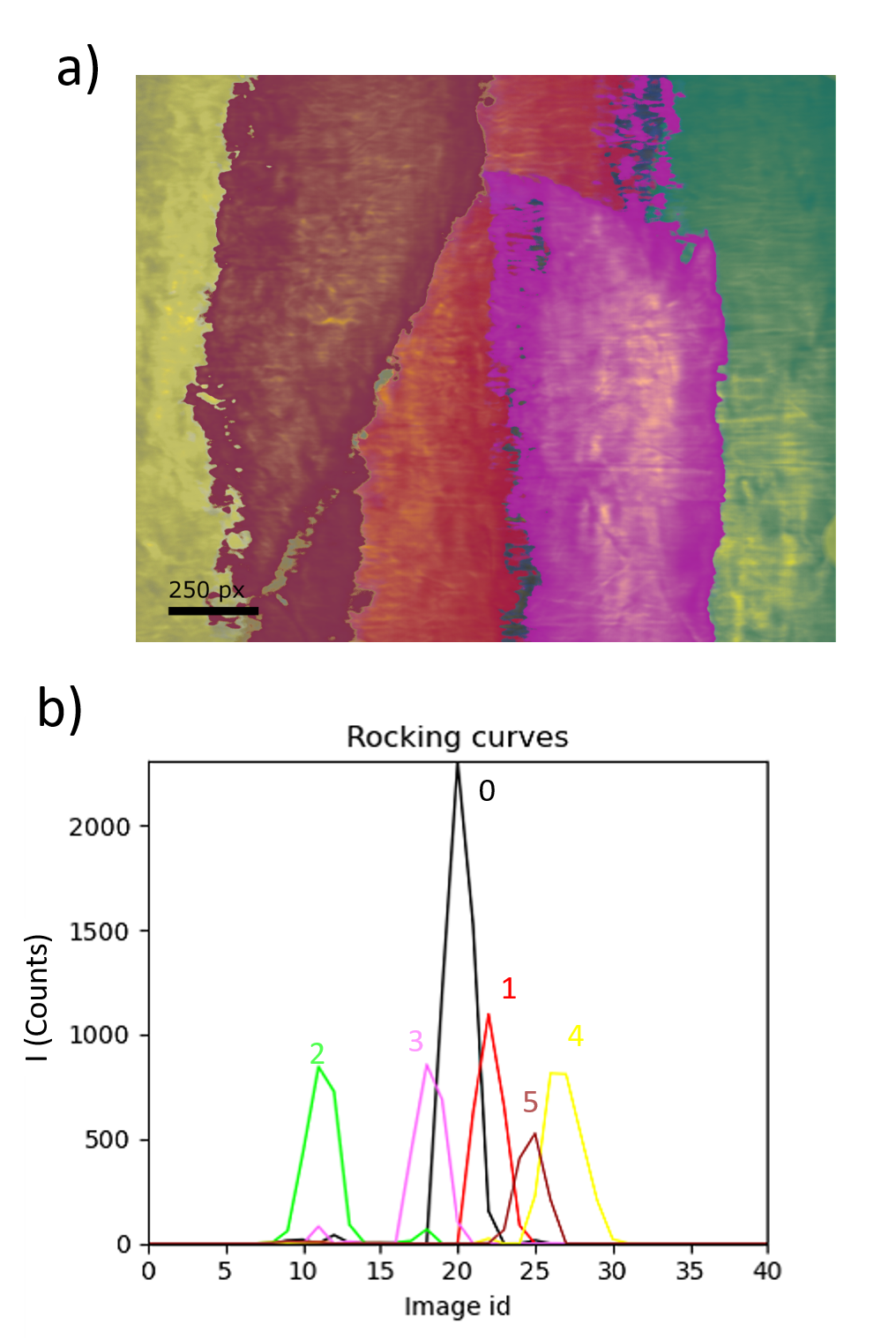}}
    \end{center}
    \caption{ Example of blind source separation of obtained from a 1D rocking scan of a (200) reflection of an Al single crystal \cite{LeoraPrivate}.a) Rocking curve image of 0.2 deg in $\mu$. Colored regions correspond to blind source separated components as shown in b) Rocking curves for each independent component.
    }
\label{figure-s4}
\end{figure}

In Fig.~\ref{figure-s4} the result of the blind source separation is shown for the rocking curve ($\mu$) obtained for a single crystal of Al in a range of 0.2 deg for a $(200)$ reflection.
In Fig.~\ref{figure-s4}a the components resulting from the separation are identified by different colors. Fig.~\ref{figure-s4}b follows the same color code to show the rocking curves corresponding to each component.
The blind source separation operation can be performed on 2D maps, the output can in that case also be plotted as an rsm for each component. 
Once the components are obtained they can be exported as HDF5 files.
%
%
\textit{darfix} offers the possibility to link components from two different datasets. 
This allows tracing features from one layer to the next in multi layer scans or the evolution of features under different external conditions. 


\section{Future evolution}

Due to its modular structure, it is easy to extend \textit{darfix} with additional functionalities and widgets, for example additional pre-processing tools such as gradient-based threshold removal for the detection of low-intensity features \cite{DresselhausMovies}.

The functionalities implemented so far are relatively low-level.
They mostly implement statistical methods and do not include analysis of the diffraction physics. 
This is an obvious field for future developments. 
In particular, we envisage to implement the transformation from angle-space to reciprocal space, including the transformation of peak shifts in mosaicity and strain scans to strain components \cite{Poulsen2021}.

Furthermore, modules could be developed for the identification of characteristic features in the sample, e.g.~isolated dislocations \cite{Jakobsen2019}. 
Development is in progress for the automatic tracking of mobile dislocations in time sequences \cite{DresselhausMovies}.
Bayesian inference methods can be used to improve the accuracy of the dislocation core position to ${\approx}5\un{nm}$ \cite{BrennanBayesian}.
Due to the long range of strain fields from dislocations, these techniques are directly applicable to classical diffraction topography and rocking curve imaging.

Fine intra-granular defect features of dislocation arrangements can be highlighted by plotting the local orientation gradient of the tilt angles $\mu$ and $\chi$ for neighboring voxels within each layer using the following relation: $\Delta\gamma=\sqrt{(\Delta\mu)^{2}+(\Delta\chi)^{2} }$ \cite{Mavrikakis2019, Ahl2017}. Here, $\Delta\mu$ and $\Delta\chi$ are the differences between the local sample tilt COM and their grain averages. This is achieved by taking the spatial derivatives of the center-of-mass maps, providing information on the dislocation density \cite{pantleon2008resolving, simons2019nondestructive}. 

Another field for future development is the transformation of a series of 2D scans into a 3D volume model of the sample. 
At present this is done manually for layer scans \cite{Yildirim2022}, but stacking and registration could be automated.
An alternative measurement strategy would be topo-tomo scans \cite{Ludwig2009}, where projections are recorded while the sample is rotated about the scattering vector.
Results should be saved in a format compatible with dedicated 3D analysis software such as Paraview \cite{Paraview1,Paraview2}.
 Work for the 3D segmentation of dislocation networks is ongoing \cite{DresselhausNetworks}.



\section{Conclusions} 

\textit{darfix} is a Python package for the analysis of dark-field x-ray microscopy and diffraction topography data.
It provides data processing and visualization tools that can be used either as library components or via a graphical user interface as an add-on to an Orange workflow.

\textit{darfix} includes analysis functions specific to common scan types used in DFXM, such as rocking curve imaging, mosaicity and strain scans.

Through blind source separation, different diffracting elements present on the same map or rocking curve can be identified as distinct sources of diffraction, associating each separated unit to their corresponding rocking curve and reciprocal space map.













\ack{Acknowledgements}

We acknowledge the European Synchrotron Radiation Facility for provision of synchrotron radiation facilities and we would like to thank Thomas Dufrane for assistance in using beamline ID06-HXM. 
We thank Leora Dresselhaus-Marais for providing the unpublished data shown in Fig.~\ref{figure-s5} and critical reading of the manuscript. 
We thank Thu Nhi Tran Caliste for stimulating discussions on rocking curve imaging.
Sonja R.~Ahl and Hugh Simons contributed the original data analysis scripts for mosaicity maps that are now implemented in \textit{darfix}.

\referencelist[darfix]

\end{document}